# AISC deployment in dynamic UAV-assisted MEC network: a reinforcement learning method based on heterogeneous graph attention neural network

Hanzhi Chang, Jing Bai, Xin Tang, Xiaomei Liu

*Abstract*—**Unmanned aerial vehicles-assisted mobile edge computing (UMEC) can execute compute-intensive and latency-critical artificial intelligence (AI) services, which can be provided by multiple UAVs collaborating in the air to perform inference tasks. Completing an AI service requires multiple inferences, each of which is implemented by an AI service chain consisting of multiple virtual network functions (VNFs). The application of AISC relies on an efficient AISC deployment strategy to determine which UAV to deploy VNF on. However, the UMEC network topology is highly dynamic due to the high-speed movement of UAVs or their departure/arrival, which makes the AISC deployment in the UMEC network challenging. In addition, the intricate relationships between UMEC environment and AISC, as well as between individual VNFs in an AISC, can also affect the effectiveness of AISC deployment strategy. Moreover, under the constraints of energy consumption and load balancing, it is also difficult to optimize the AISC strategy to minimize AISC completion time for enhancing the quality of AI service. To address the above challenges, this paper proposes a double deep attention Q-network based on heterogeneous graph neural networks, which incorporates heterogeneous graph to capture diverse relationships in UMEC and utilizes attention mechanisms to adaptively focus on critical nodes and links for intelligent AISC deployment. The experimental results demonstrate that the proposed algorithm performs excellently in AISC completion time, AISC completion rate, load balancing and energy consumption.**

## I. INTRODUCTION

UNMANNED aerial vehicles (UAVs) have the advantages of flexible deployment, rapid response and wide coverage to make up for the lack of ground serves [1][2], enabling mobile edge computing (MEC) to execute compute-intensive and latency-critical artificial intelligence (AI) services such as face recognition, unmanned driving and augmented reality [3]. In UAV-assisted MEC (UMEC) network, multiple UAVs collaborate in the air to provide AI services for users by performing inference tasks [4]. An AI service consists of multiple inference tasks [9], each of which can be treated as an AI service chain (AISC) provided by several virtual network functions (VNFs) connected in series [5]. As a practical example, consider an AI-driven video analytics service chain deployed across multiple low-energy integrated communication units. In such a scenario, raw video frames are first preprocessed and compressed in the initial VNF. Then, object detection or feature extraction is performed in the second VNF. Finally, the last VNF conducts tasks such as activity recognition or anomaly detection. Fig. 1 shows the framework of AISC deployment. In the User Layer, users in different areas initiate AISC requests through their associated base stations. These service requests are transmitted to the UMEC Network Layer, where multiple UAVs form a flexible airborne MEC infrastructure. Each VNF in an AISC is mapped to a specific UAV for execution. As defined in [6], with the assistance of the orchestrator, each VNF is placed on an appropriate UAV and executed on its onboard computing module, which typically consists of an embedded CPU/GPU platform.

UMEC networks are increasingly used to execute latency-critical AI services such as video analytics, surveillance, and environmental monitoring, where inference tasks must be processed collaboratively across multiple UAVs. Since each VNF in an AISC relies on the computing and communication resources of its hosting UAV, the placement of VNFs directly determines the end-to-end service delay, energy efficiency, and overall service reliability. An inappropriate deployment may lead to significant performance degradation. For example, placing successive VNFs on distant or overloaded UAVs can cause excessive transmission delay or even service interruption when the UAV becomes unavailable. Therefore, designing an efficient and intelligent AISC deployment strategy is essential for ensuring stable, reliable, and high-quality AI services in UMEC networks. In order to fully utilize UAV resources and optimize the quality of AI services, an efficient AISC deployment strategy is required to decide on which UAV each VNF in an AISC is deployed. Fig. 2 (a) shows the illustration of AISC deployment in the UMEC network. However, there are the following challenges in deploying AISC in the UMEC network.

- The high-speed movement of UAVs leads to the high dynamic UMEC network topology [7]. Once the network topology changes, the current deployment may not meet user needs. The virtual links of AISC must be remapped

Corresponding author: Jing Bai
Hanzhi Chang, Jing Bai, Xin Tang and Xiaomei Liu are with the School of Cyber Science and Engineering, University of International Relations, Beijing, China. E-mail: {hanzhi.chang, baijing, xtang, liuxiaomei}@uir.edu.cn.

to new links, as shown in Fig. 2 (b). In addition, when UAVs leave the network, VNFs deployed on these UAVs must be migrated and the AISCs whose data transmission through the departing UAVs need to be redeployed, as shown in Fig. 2 (c). When new UAVs join the network, the deployment space expands, allowing subsequent AISCs to be deployed on the newly joined UAVs, as shown in Fig. 2 (d). Therefore, the real-time deployment of AISC in the dynamic UMEC network is a problem that needs to be solved.

- AISC completion time can be used to effectively characterize the quality of AI service. However, minimizing AISC completion time and optimizing load balancing restrict each other. In addition, AISC completion time is also affected by energy consumption. If the UAV leaves the network due to energy exhaustion, VNFs deployed on it must be migrated, which increases AISC completion time [8]. Therefore, how to minimize AISC completion time while optimizing load balancing and energy consumption is a problem that needs to be solved.
- Different from individual services, the deployment of AISC composed of multiple VNFs in the UMEC network is affected by the correlation relationships between UAVs, between AISCs and VNFs, between VNFs, between VNFs and UAVs, and between virtual links connecting VNFs and U2U links. Therefore, how to incorporate these correlation relationships into the AISC deployment process to improve the performance of the algorithm is a problem that need to be solved.

However, existing studies either assumed that the network topology is static [15][18][19][20][25], or focused on the deployment of individual service in the UMEC network [11][23][24], or optimized only one or two of the completion time, energy consumption and load balancing [12][14][16][17][20]. To address the above challenges, this paper proposes a double deep attention Q-network based on heterogeneous graph neural networks (DDAQ-HGNN). To the best of our knowledge, this is the first work to incorporate heterogeneous graph modeling into the AISC deployment process. The main contributions of this paper are summarized as follows:

- We model AISC deployment problem in the dynamic UMEC network as an Integer Linear Programming (ILP) problem to jointly optimize three critical objectives: energy consumption, AISC completion time and network load balancing. In particular, we propose a migration model for capturing the system behaviors in scenarios where the network topology changes over time and UAVs leave/join the network.
- The proposed DDAQ-HGNN algorithm consists of three components: heterogeneous graph encoder, heterogeneous attention Q-network and heterogeneous nodes decoder. The encoder integrates all correlation relationships in the UMEC network into a unified heterogenous graph representation. The Q-network incorporates a heterogeneous graph attention mechanism to effectively extract features from multi-type nodes and edges. The decoder predicts UAVs selection based on the features learned from the heterogeneous graph, enabling decision-making under dynamic environments.
- To enable DDAQ-HGNN to better evaluate long-term deployment performance, we design an asynchronous reward function. That is, the agent does not receive immediate rewards upon each VNF deployment, but calculates and returns rewards after the entire AISC completes its inference process or fails, thereby guiding the algorithm to make more informed decisions.
- We design and conduct extensive experiments to evaluate the adaptability and performance of the proposed algorithm in the dynamic UMEC network. We verify the performance of the algorithm under different network topologies. Additionally, we assess the adaptability of the algorithm under varying UMEC networks topology change probabilities. The results demonstrate that our proposed algorithm significantly outperforms existing baseline algorithms in terms of AISC completion time, network load balancing, AISC completion rate and energy consumption.

The rest of this paper is organized as follows. Section II discusses the related work. Section III introduces the system model and problem formulation. Section IV presents the proposed DDAQ-HGNN in detail. Section V shows the results of performance evaluations. Finally, we conclude the paper and discuss future research directions in Section VI.

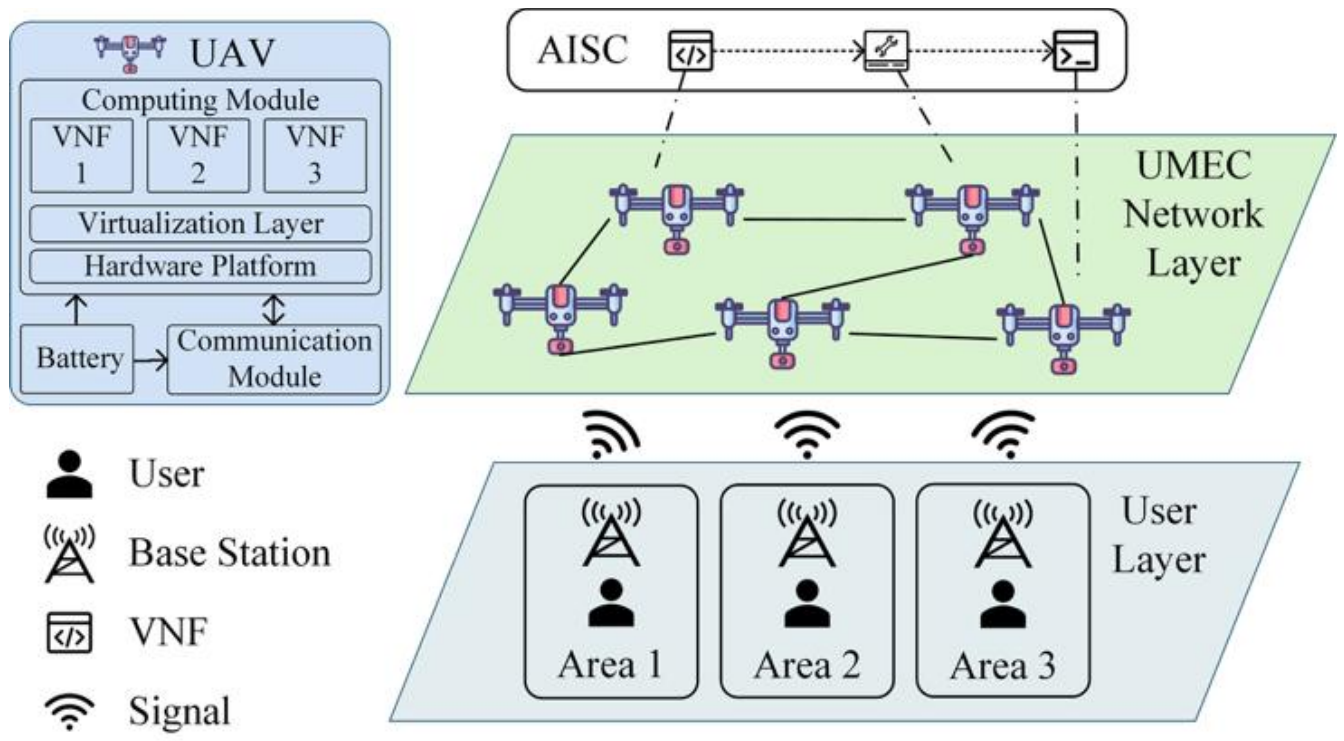

**Fig. 1.** The framework of AISC deployment

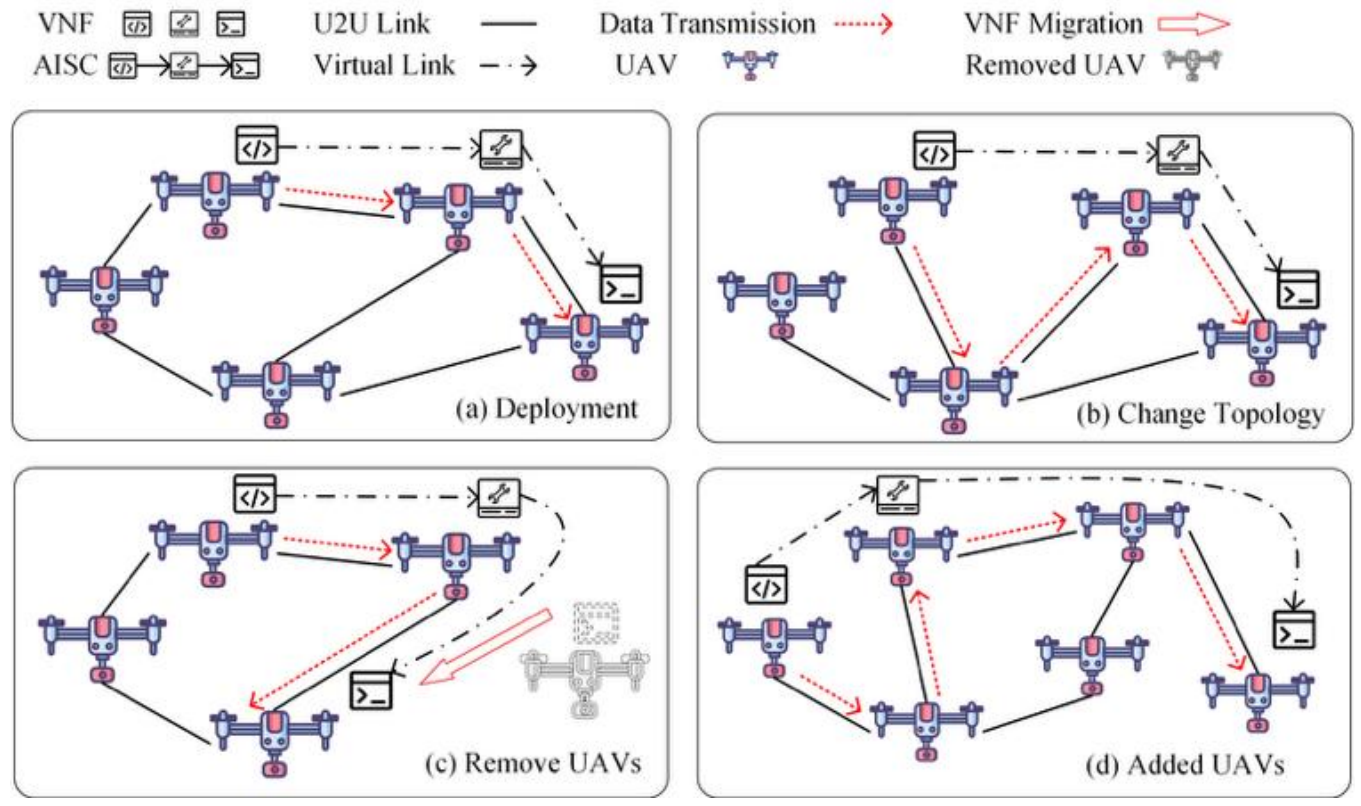

**Fig. 2.** Illustration of AISC deployment in the dynamic UMEC network

## II. Related Works

In this section, we introduce the existing studies related to our work from three aspects: RL for service deployment and scheduling, GNN for network modeling and scheduling and multi-objective optimization in UMEC networks.

### *A. RL for Service Deployment and Scheduling*

Currently, solutions for service placement and scheduling are generally divided into two categories: heuristic solutions and RL solutions. Liu *et al*. [12] considered cost-oriented and delay-constrained anycasting in space–air–ground networks as a mixed-integer program refined by a greedy search, assuming quasi-static links and fixed number of nodes. Xu *et al.* [14] proposed a cloud-centric heuristic method for mapping UAV-sourced video service chain requests, which minimizes network cost but depends on centrally controlled and stable infrastructures. Wang *et al*. [16] proposed a two-stage offline scheduler, ToRu, targeting heterogeneous multi-UAV edge computing, which first deploys the chains in parallel when resources are abundant and then serially reorders tasks to maximize revenue under CPU/FPGA constraints. Wang *et al*. [17] proposed embedding and migration algorithms for UAV IoT scenarios, specifically addressing the challenge of maintaining service continuity under UAV mobility and changing IoT demands, with the main objective of reducing service latency and migration overhead. Because these heuristic solutions lack the ability to dynamically learn from real-time environmental feedback, it is often difficult for them to effectively adapt to the rapid and frequent topology changes in dynamic UMEC networks [12][14][16][17].

RL has become a powerful solution for service deployment and scheduling, owing to its advantages in adaptability, real-time decision-making, and the capability to learn effective policies from environmental interactions without relying on explicit network models. Wu *et al.* [15] developed a deep reinforcement learning (DRL)-based QoE-aware service chain deployment framework for UAV networks, focusing on improving service quality by dynamically adjusting service chain configurations. Zhang *et al*. [19] proposed a mobility-aware service chain deployment strategy in network function virtualization (NFV)-based edge-cloud environments to proactively reduce service downtime caused by UAV mobility using predictive migration techniques. Additionally, Wang *et al*. [20] proposed a dual-agent DRL framework for MEC scenarios, which jointly optimizes partial task offloading and service chain mapping to effectively reduce overall latency and improve resource utilization efficiency. Bao *et al*. [25] applied multi-agent DRL with action masking to achieve sustainable task offloading in UAV-assisted smart farm networks, focusing on secure and efficient resource management. The methods in [15][20][25] assumed that the network is static and were not suitable for analyzing the dynamic UMEC network, where the topology and the number of UAVs fluctuate frequently. Moreover, the methods in [18][19][20] optimized only one or two of the completion time, energy consumption and load balancing, ignoring the synergistic impact of these metrics on AISC deployment strategy.

### *B. GNN for Network Modeling and Scheduling*

GNNs have become a powerful tool for network resource management and service placement, because they are able to learn from graph-structured data to capture the relational and topological characteristics of networks. Ping *et al*. [22] proposed a GNN-based solution for task scheduling-dependent QoE optimization in edge-cloud computing networks, demonstrating the benefits of graph-structured modeling in task coordination and resource management. Feng *et al*. [24] utilized graph attention mechanisms combined with RL to design UAV trajectories and allocate communication resources. Gao *et al*. [23] integrated federated RL with GNNs to address trajectory optimization in UAV networks. However, most of these solutions are designed for relatively static scenarios, where the network topology and the number of nodes remain stable over time [22][23][24]. Such settings simplify the modeling process but limits generalization to more dynamic environments. In contrast, deploying AISC in UMEC networks introduces unique challenges: UAVs may join or leave the network dynamically, inter-node relationships evolve with mobility, and the overall network topology changes over time. Additionally, since AISC deployment involves placing VNFs one by one, it is necessary to consider sequential decision dependencies. These characteristics require adaptive models that can capture dynamic topological changes and heterogeneous relationships. Therefore, non-heterogeneous graph GNNs are not suitable for supporting the adaptive and real-time decision-making required for AISC deployment in UMEC networks.

### *C. AISC Deployment in MEC networks*

There are fewer studies on the AISC deployment approach in UMEC networks. Lu *et al.* [18] proposed a RL-based framework for dynamically deploying service chain in UAV networks under security threats, optimizing deployment success rate and security robustness by intelligently avoiding vulnerable network paths. However, this study only considered deployment success rate and cost, without incorporating delay and load balancing, both of which are critical for ensuring timely end-to-end service execution and preventing resource congestion in practical UMEC environments. Wang *et al*. [21] proposed a distributed generative reinforcement learning proposed a distributed generative RL framework to achieve stable service chain orchestration in highly dynamic UAV swarm networks. Their method leveraged generative models to predict future network states, thereby supporting coordinated decision-making among distributed UAVs, which improves orchestration stability and reduces the impact of network fluctuations. However, this study only considered the deployment problem in static UMEC networks and did not take into account scenarios in which the number of UAVs changes or the UAV network topology evolves. These factors are crucial in real-world UMEC systems, because UAV arrivals, departures, and topology variations can significantly affect link availability, resource allocation, and end-to-end service performance. Qiu et al. [3] proposed an online security-aware and reliability-guaranteed provisioning scheme for AISCs in edge intelligence clouds. Their model integrated security constraints and reliability requirements into a stochastic optimization framework to improve service stability. However, this approach relied on heuristic algorithm, leading to its limited adaptability to dynamic network states. Moreover, in terms of optimization objectives, it only focused on minimizing cost and improving security levels without considering delay performance. Li *et al*. [13] presented a slicing-based AISC provisioning method that balanced AI service performance and data-management resource consumption. However, this study relied on static edge servers as the execution platforms for VNFs and did not consider UAVs as MEC service nodes, ignoring the flexibility and mobility advantages offered by UAV-assisted MEC systems. In addition, the method only optimized the cost, ignoring other essential performance indicators such as end-to-end delay, load balancing, and service success rate. These factors are critical in practical UMEC environments, because they directly influence user experience, system stability, and the feasibility of sustained AI service deployment. Moreover, none of these studies considered the problem of service migration when nodes fail.

In contrast, we address the limitations identified above. Specifically, we employ a heterogeneous graph neural network to support AISC deployment when both the number of UAV and the UMEC network topology change over time. In addition, our model explicitly incorporates VNF migration to maintain service continuity in dynamic UMEC networks. By integrating attention mechanisms with reinforcement learning, the proposed approach achieves joint optimization of energy consumption, AISC completion time, and network load balancing.

## III. System Model and Problem Formulation

In this section, we propose a system model for AISC deployment in the dynamic UMEC network, which includes the physical model, the migration model, and the computation model. Based on the system model, we propose the problem constraints and formulate the optimization problem as an ILP problem.

### A. Physical Model

At time slot $t$, the UMEC network is represented as a connected graph $G_t = \{U_t, E_t\}$, where $U_t$ denotes the set of UAVs and $E_t$ denotes the set of U2U links. We use $|U_t|$ and $|E_t|$ to denote the total number of UAVs and U2U links, respectively. Each UAV is denoted as $u_{t,i} \in U_t (i \in [1, |U_t|])$. Each U2U link is denoted as $e_{t,j} \in E_t$, representing an undirected U2U link between two UAVs $u_{t,i}$ and $u_{t,i'}$ $(i' \in [1, |U_t|], i \neq i', j \in [1, |E_t|])$. The attributes of the UAVs and U2U links are listed in Table I.

Each AISC is represented as a directed graph $\mathcal{F}_f = \{\mathcal{V}_f, \mathcal{L}_f\}$, where $f$ denotes the index of the AISC. $\mathcal{V}_f$ is the set of VNFs, and $\mathcal{L}_f$ is the set of virtual links. We use $|\mathcal{V}_f|$ and $|\mathcal{L}_f|$ to denote the number of VNFs and virtual links, respectively. Each VNF is denoted as $v_{f,m} \in \mathcal{V}_f$ $(m \in [1, |\mathcal{V}_f|])$. A virtual link is represented as $l_{f,n} \in \mathcal{L}_f$, indicating a directed virtual link between two connected VNF $v_{f,m}$ and $v_{f,m'} \in \mathcal{V}^f$ $(m' \in [1, |\mathcal{V}_f|], m \neq m', n \in [1, |\mathcal{L}_f|])$. The attributes of VNFs and virtual links are listed in Table II.

### B. Problem Constraints

At any time slot $t$, the deployment and migration of AISCs must satisfy the following constraints:

**Mapping Constraints.** At each time slot $t$, each VNF can only be deployed on one UAV, while each UAV can host multiple VNFs. This constraint is denoted as:

$$\sum_{i=1}^{|U_t|} \lambda_{t,f,m,i} = 1, \sum_{f} \sum_{m=1}^{|\mathcal{V}_f|} \lambda_{t,f,m,i} \geq 0 \tag{1}$$

where $\lambda_{t,f,m,i} \in \{0,1\}$, and $\lambda_{t,f,m,i} = 1$ indicates that VNF $v_{f,m}$ is deployed on UAV $u_{t,i}$ at time slot $t$; otherwise, it is not deployed. At each time slot $t$, any virtual link can be mapped to multiple U2U links, and each U2U link can carry multiple virtual links. This constraint is denoted as:

$$\sum_{j=1}^{|E_t|} \lambda_{t,f,n,j} \geq 1, \sum_{f} \sum_{n=1}^{|\mathcal{L}^f|} \lambda_{t,f,n,j} \geq 0 \tag{2}$$

where $\lambda_{t,f,n,j} \in \{0,1\}$, and $\lambda_{t,f,n,j} = 1$ indicates that virtual link $l_{f,n}$ is deployed on U2U link $e_{t,j}$ at time slot $t$; otherwise, it is not deployed.

**Capacity Constraints.** At each time slot $t$, every UAV and U2U link must satisfy capacity constraints. For UAVs, the constraints include computing unit capacity, memory capacity, and energy availability, which are denoted as:

$$\sum_{f} \sum_{m=1}^{|\mathcal{V}_f|} \lambda_{t,f,m,i} * c_{f,m}^{\text{num}} \leq c_{t,i}^{\text{capnum}} \tag{3}$$

$$\sum_{f} \sum_{m=1}^{|\mathcal{V}_f|} \lambda_{t,f,m,i} * c_{f,m}^{\text{mem}} \leq c_{t,i}^{\text{capmem}} \tag{4}$$

$$c^{\text{capeng}}(u_i^t) > 0 \tag{5}$$

For U2U links, the bandwidth constraint is given by:

$$\sum_{f} \sum_{n=1}^{|\mathcal{L}_f|} \lambda_{t,f,n,j} * c_{f,n}^{\text{bw}} \leq c_{t,j}^{\text{capbw}} \tag{6}$$

**AISC Completion Time Constraints.** To ensure service quality, the transmission delay of each AISC must not exceed its maximum delay requirement, which is denoted as:

$$\text{time}_f^{\text{delay}} \leq c_f^{\text{delay}} \tag{7}$$

Each AISC is allowed to occupy network resources only when its inference step does not meet the requirement, which is denoted as:

$$c_f^{\text{stp}} < c_f^{\text{reqstp}} \tag{8}$$

## IV. Solution

To address the challenges posed by multi-dimensional states, complex topologies, and multi-objective optimization in the dynamic UMEC network for AISC deployment, this paper proposes a novel RL algorithm—double deep attention Q-network based on heterogeneous graph neural networks (DDAQ-HGNN). By constructing the UMEC network as a heterogeneous graph, integrating attention mechanisms with Q-learning strategies, and predicting UAV selection based on the learned heterogeneous representations, the algorithm achieves efficient AISC deployment in dynamic UMEC networks. The entire algorithm can be formalized as an MDP, as illustrated in Fig. 3.

In the MDP framework, DDAQ-HGNN acts as the agent that captures observations from the environment, executes actions, and receives feedback in the form of rewards. In each interaction, the environment provides an observation that consists of the current UMEC network topology, node and edge attributes, and pending AISC deployment.

DDAQ-HGNN agent consists of three main components. The heterogeneous graph encoder encodes the input observation into a unified graph representation, preserving type-specific information and contextual dependencies within the UMEC network to provide a structured foundation for decision-making. The heterogeneous attention Q-network applies a graph attention mechanism to perform cross-type information fusion and importance modeling for different types of nodes and edges. This process generates high-dimensional feature vectors representing candidate deployment nodes (i.e., UAVs). This design enables the model to capture the joint impact of local attributes and global topology dynamics on deployment decisions, thereby improving the generalization and robustness of the learned policy. The heterogeneous nodes decoder maps each node embedding to a scalar Q-value and selects the optimal UAV node for the current VNF deployment based on the highest Q-value.

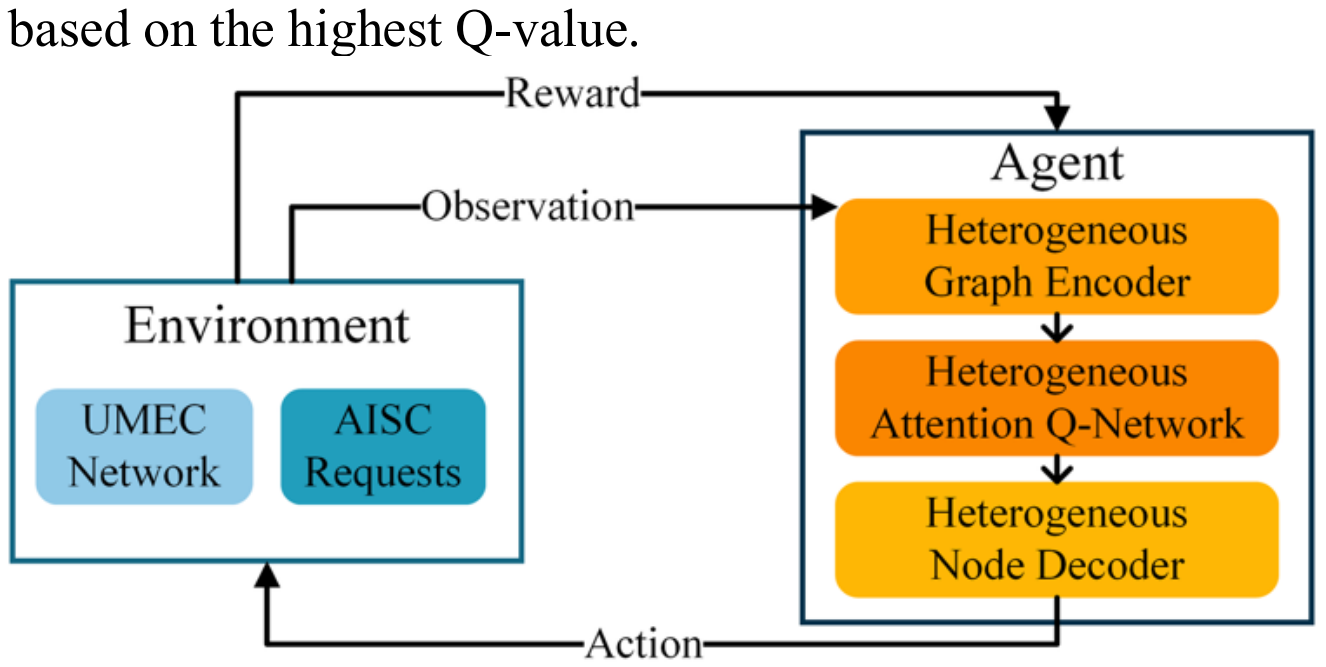

**Fig. 3.** The MDP framework of DDAQ-HGNN for AISC deployment in dynamic UMEC networks

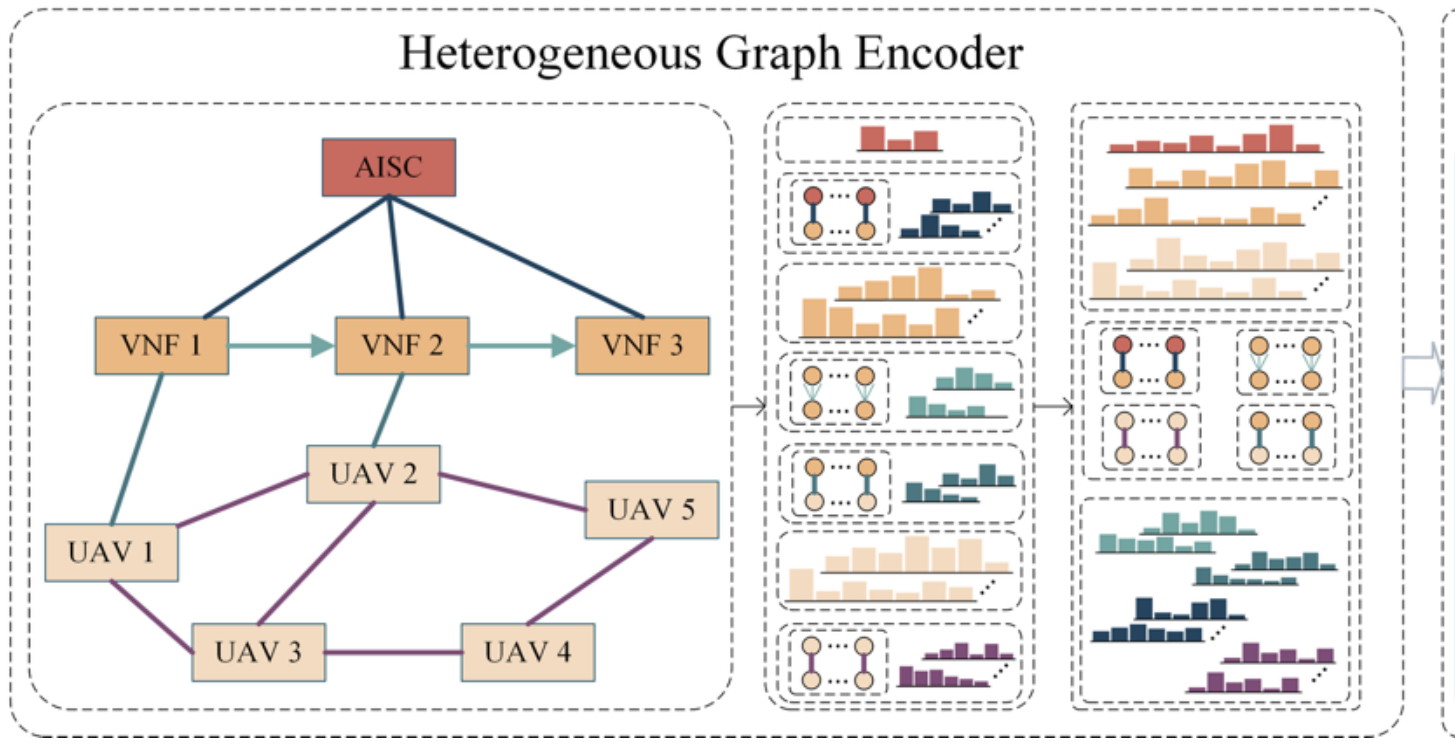

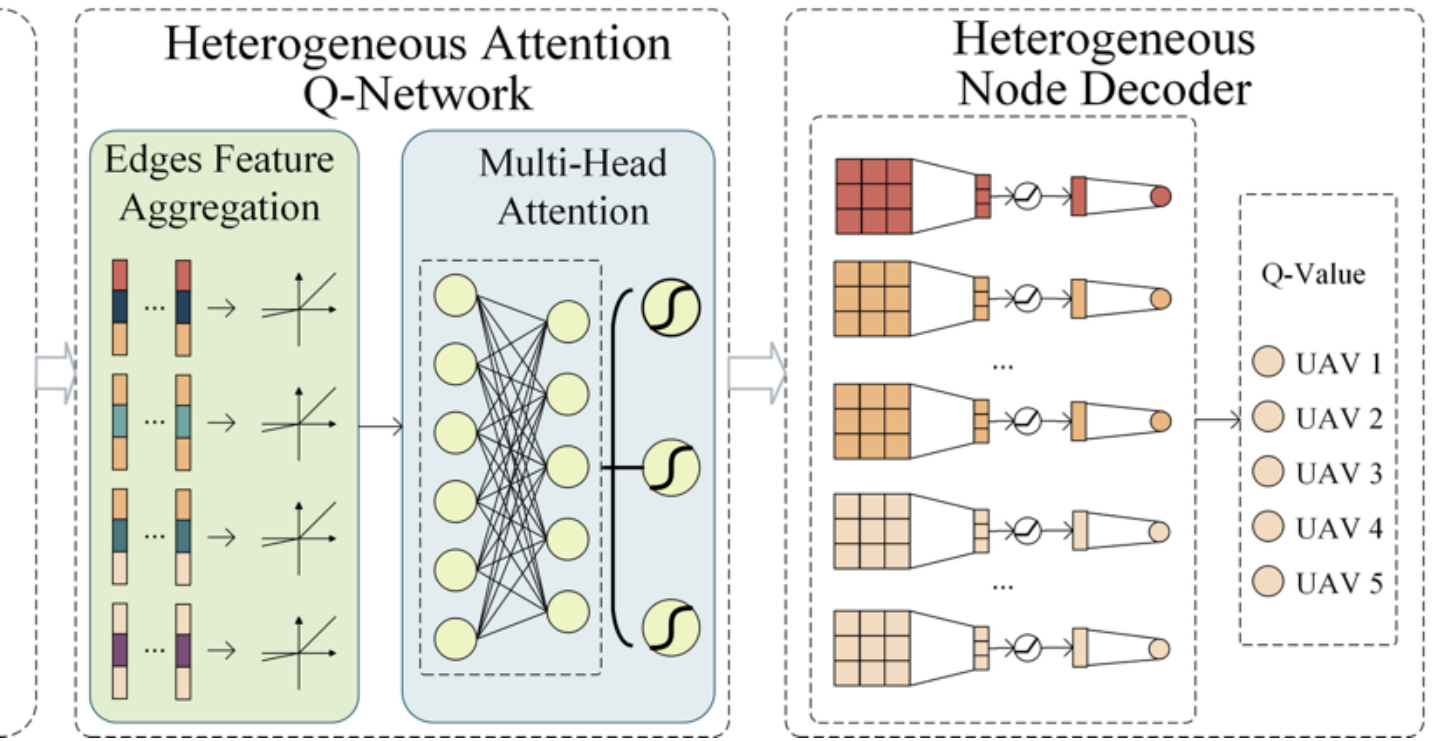

**Fig. 4.** Detailed architecture of DDAQ-HGNN

The environment then provides rewards based on the effectiveness of the AISC deployment action, considering factors such as energy consumption, network load balancing, and resource conflicts to guide policy updates.

In addition, we design a custom asynchronous reward function to support multi-objective optimization during the AISC deployment process. This reward function simultaneously accounts for key performance metrics including energy consumption, average AISC completion time, network load balancing, and AISC completion rate. Through continuous interaction between the RL agent and the dynamic environment, DDAQ-HGNN progressively converges to a stable and efficient deployment policy under variable topologies and uncertain resource conditions.

The rest of this section will introduce the three components of DDAQ-HGNN and training algorithm in detail based on Fig. 4.

**Remark.** The UMEC environment features highly dynamic network topologies due to the continuous movement of UAVs. It also contains heterogeneous structures composed of multiple node types, such as UAVs and VNFs, and multiple edge types with attributes like bandwidth and delay that directly affect deployment decisions. To effectively capture these characteristics, we adopt HGNN, which supports type-specific message passing and incorporates edge features, enabling a more expressive modeling of the UMEC network than homogeneous GNNs. Furthermore, DDAQ leverages attention mechanisms to adaptively emphasize the most informative nodes and links under dynamic conditions, allowing the agent to learn robust deployment policies. Therefore, DDAQ-HGNN address the challenges of heterogeneous relationships, topology variability, and multi-objective optimization in AISC deployment.

## *A. Heterogeneous Attention Q-Network*

The heterogeneous attention Q-network employs the graph attention mechanism to fuse the features of different types of nodes and edges to produce high-dimensional feature vectors representing each candidate UAV.

### 1) Edges Feature Aggregation

For each edge in the heterogeneous graph, the edge attribute and edge type embeddings are concatenated to form the edge feature, denoted as $\vec{e}_{xy} = [\vec{d}_{xy}^{\text{eattr}} || \vec{d}_{xy}^{\text{type}}]$ . $\vec{d}_{xy}^{\text{eattr}}$ is the corresponding row vector from $\mathbf{s}_t^{\text{eattr}}$ , and $\vec{d}_{xy}^{\text{type}}$ is the corresponding column vector from $\mathbf{s}_t^{\text{type}}$ . The symbol $xy$ indicates that the edge starts from node $x$ and ends at node $y$ ($x, y \in \mathcal{N}_t$). For each destination node $y$, the edge feature $\vec{e}_{xy}^{+} = [\vec{s}_y^t || \vec{e}_{xy}]$is combined with the node's own attribute to form an enriched input representation for the attention mechanism. $\vec{s}_{t,y}$ is the corresponding row vector of node $y$ in $\mathbf{s}_t^{\text{nattr}}$. The vector $\vec{e}_{xy}^{+}$ is then used to compute the attention coefficient $\alpha_{xy}$ by the shared mechanism. The computation is defined as follows:

$$\alpha_{xy} = \frac{\exp\left(\text{LeakyReLU}((\vec{w}^{\text{att}})^{\text{T}} \cdot [\vec{s}_{t,x} || \vec{e}^{+}_{xy}])\right)}{\sum_{z\in\mathcal{N}_{t,x}} \exp\left(\text{LeakyReLU}((\vec{w}^{\text{att}})^{\text{T}} \cdot [\vec{s}_{t,x} || \vec{e}^{+}_{xz}])\right)} \quad (9)$$

where $\mathcal{N}_{t,x}$ denotes the set of neighboring nodes of node $x$ in the heterogeneous graph and $\vec{s}_{t,x}$ is the corresponding row vector for node $x$ in $\mathbf{s}_t^{\text{nattr}}$. A shared single-layer perception network $\vec{w}^{\text{att}}$ of size $2 + k^{\mathcal{N}} + k^{\text{type}}$ is used to compute attention scores. The attention coefficient $\alpha_{xy}$ reflects the importance of node $y$ to node $x$, considering both node and edge features. The LeakyReLU activation function is defined as:

$$\text{LeakyReLU}(p) = \begin{cases} p & p > 0 \\ \beta p & p \le 0 \end{cases} \quad (10)$$

where $\beta \in \mathbb{R}$ is the leakage factor.

**2) Multi-Head Attention**

The attention coefficients are used to update the feature representation of each node. To incorporate both edge attributes and node attributes, a weighted fusion is performed to compute the updated node feature $(\vec{s}_{t,x})'$:

$$(\vec{s}_{t,x})' = \sigma\left(\sum_{y\in\mathcal{N}_{t,x}} \alpha_{xy}\, \vec{w}^{\text{node}} [\vec{d}^{\text{eattr}}_{xy} || \vec{s}_{t,x}]\right) \quad (11)$$

where $\vec{w}^{\text{node}} \in \mathbb{R}^{(k_{\mathcal{N}} + k_{\mathcal{E}})}$ is the perception vector applied to connect edge attribute and node feature. The edge type embedding $\mathbf{s}_t^{\text{type}}$ is excluded from this fusion because it is a discrete feature and has already been incorporated in the attention coefficient calculation. The activation function $\sigma(\cdot)$ is defined as:

$$\sigma(p) = \frac{1}{1 + \theta^{-p}} \quad (12)$$

where $\theta$ denotes the natural base and is used here to avoid notation conflicts.

Finally, in order to stabilize the self-attention mechanism, the heterogeneous attention Q-network employs multiple independent attention heads. The final node feature is obtained by concatenating the outputs of each attention head:

$$(\vec{s}_{t,x})' = \left[\sigma\left(\sum_{y\in\mathcal{N}_{t,x}} \alpha^{\kappa}_{xy}\, \vec{w}^{\kappa}_{\text{node}} [\vec{d}^{\text{eattr}}_{xy} || \vec{s}_{t,x}]\right)\right]^{\text{K}}_{\kappa} \quad (13)$$

where K is the number of attention heads, and $\kappa$ indexes each head.

## *B. Heterogeneous Node Decoder*

The heterogeneous node decoder maps each node feature to a scalar Q-value and selects the optimal UAV for the current VNF deployment based on these values.

The decoder first aggregates the updated node features into a matrix $\mathbf{P}_t \in \mathbb{R}^{|\mathcal{N}_t| \times k^{\text{out}}}$, which is calculated as:

$$\mathbf{P}_t = \begin{bmatrix} \left((\vec{s}_{t,x=1})'\right)^{\text{T}} \\ \cdots \\ \left((\vec{s}_{t,x=|\mathcal{N}^t|})'\right)^{\text{T}} \end{bmatrix} \quad (14)$$

Next, the matrix $\mathbf{P}_t$ is decoded into a matrix $\mathbf{P}_t^{\text{val}}$ by using a value perception network, as follows:

$$\mathbf{P}_t^{\text{val}} = \text{ReLU}(\mathbf{P}_t \mathbf{W}^{\text{val}} + \mathbf{b}^{\text{val}}) \quad (15)$$

where $\mathbf{W}^{\text{val}} \in \mathbb{R}^{k^{\text{out}} \times k^{\text{val}}}$ is the value perception matrix, and $\mathbf{b}^{\text{val}} \in \mathbb{R}^{|\mathcal{N}_t| \times k^{\text{val}}}$ is the bias matrix. The ReLU activation function is defined as:

$$\text{ReLU}(p) = \max(0, p) \quad (16)$$

The value matrix is projected into a value vector $\vec{q}_t$, which represents the Q-value for each node:

$$\vec{q}_t = \mathbf{P}_t^{\text{val}} \vec{w}^{\text{val}} \quad (17)$$

where $\vec{w}^{\text{val}} \in \mathbb{R}^{k^{\text{val}}}$ is a perception vector used for value projection.

Therefore, the final output of the agent is defined as:

$$a_t = \underset{q\in\vec{q}_t^{\text{UAV}}}{\text{argmax}}\, \vec{q}_t \quad (18)$$

where $\vec{q}_t^{\text{UAV}}$ denotes the subset of $\vec{q}_t$ corresponding to UAV nodes.

**Remark.** All trainable parameter matrices and vectors ( $\mathbf{W}^{\text{UAV}}$, $\vec{b}_i$, $\mathbf{W}^{\text{AISC}}$, $\vec{b}_f$, $\mathbf{W}^{\text{VNF}}$, $\vec{b}_m$, $\mathbf{W}^{\text{plink}}$, $\vec{b}_j$, $\mathbf{W}^{\text{vlink}}$, $\vec{b}_n$, $\mathbf{W}^{\text{type}}$, $\vec{w}^{\text{att}}$ $\vec{w}^{\text{node}}$, $\mathbf{W}^{\text{val}}$, $\mathbf{b}^{\text{val}}$, $\vec{w}^{\text{val}}$) depend only on the latent dimensions $k^{\mathcal{N}}$, $k^{\mathcal{E}}$, $k^{\text{type}}$, $k^{\text{out}}$, and $k^{\text{val}}$ and are independent of the number of nodes and edges in the heterogeneous graph. Therefore, DDAQ-HGNN is able to handle AISC deployment in dynamic UMEC networks.

# V. Simulation and Performance Evaluations

In this section, we present extensive experimental results to validate the performance of the proposed DDAQ-HGNN for deploying AISC in dynamic UMEC networks. All experiments are implemented using PyTorch [26] and NetworkX on a laptop equipped with an AMD Ryzen 7745HX CPU, 64 GB RAM, and an RTX 4060 GPU with 8 GB of memory.

## *A. Evaluation Results and Analysis*

Fig. 5 presents the average AISC completion rate achieved by different algorithms for four network topologies. DDAQ-HGNN achieves the highest completion rate under all topologies and topology change probabilities, demonstrating its superior deployment stability and task execution capability in dynamic network environments. It is worth noting that even in the extreme condition where the topology change probability reaches 30%, DDAQ-HGNN maintains a significantly higher completion rate than other algorithms, highlighting its robustness and generalization ability. As the topology change probability increases, the AISC completion rate of all algorithms decrease, indicating that dynamic topological changes have a negative impact on the deployment success rate. However, DDAQ-HGNN has the smallest performance degradation, indicating that it has a stronger adaptability to network topology perturbations. PPO and HANConv rank second in AISC completion rate, but their stability declines under high topology change probability, particularly in the Watts-Strogatz and Random Regular topologies. Their completion rates drop rapidly, revealing their limited ability to handle frequent structural changes.

Fig. 6 illustrates the average reward of different algorithms for four networks. Fig. 6(a) shows the average reward of the six algorithms under different training episodes. The proposed DDAQ-HGNN achieves the highest average reward in the later stages of training, indicating its superior long-term performance. In the early stage, DDAQ-HGNN also demonstrates a competitive convergence speed second only to PPO, reflecting its strong training efficiency. Moreover, algorithms that leverage heterogeneous graph structures generally outperform those that do not, indicating that heterogeneous graph modeling significantly enhances representational capacity and decision-making performance when addressing AISC deployment in the dynamic UMEC network. Fig. 6(b) shows the average rewards of each algorithm under four different network topologies. DDAQ-HGNN consistently achieves the highest average reward under four topologies, demonstrating its strong generalization ability to diverse network structures. In contrast, algorithms such as PPO, HANConv, and GCN show more diverse performance depending on the topology. Traditional strategies such as Greedy and Random perform poorly under four scenarios, indicating that they cannot adapt to the dynamic demands of AISC deployment in the UMEC network.

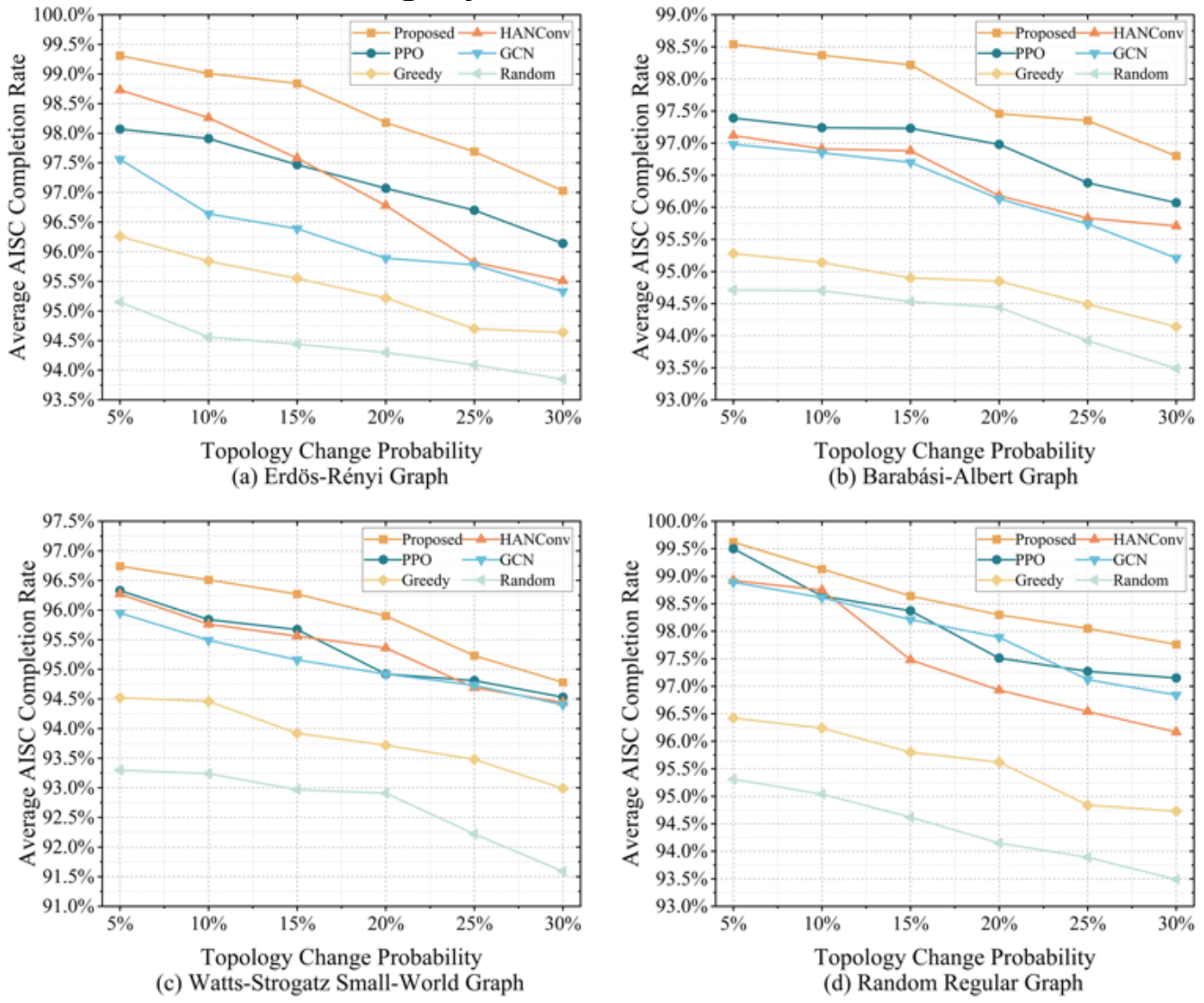

**Fig. 5.** Average AISC completion rate of different algorithms under varying topology change probabilities for four network topologies

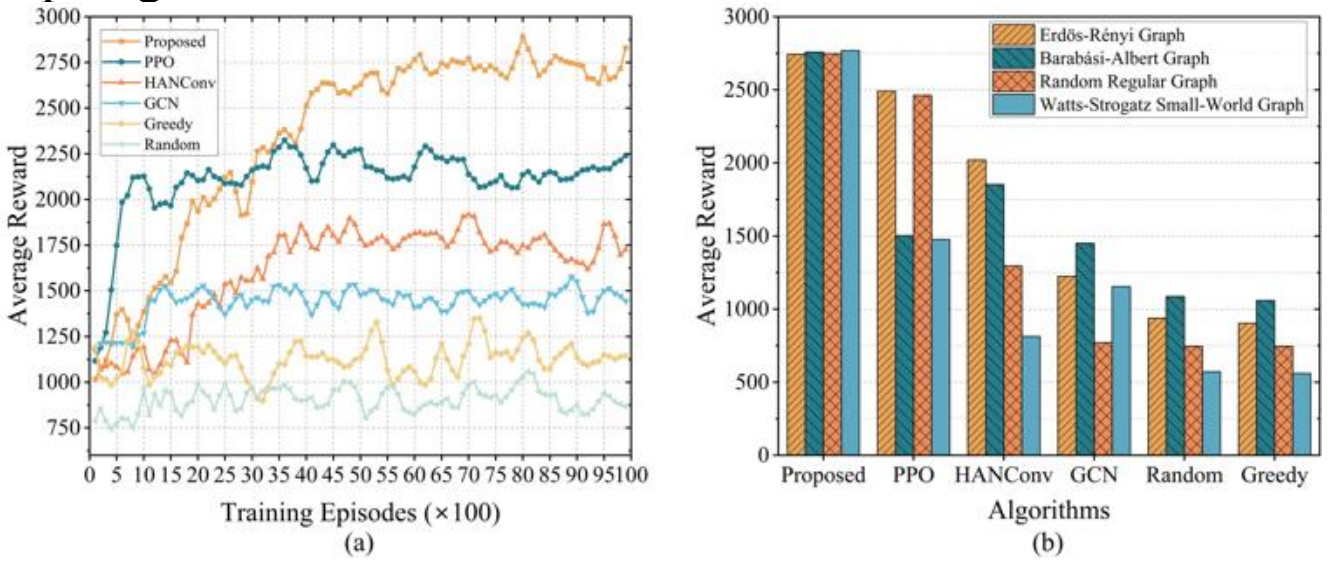

**Fig. 6.** Average reward of different algorithms for four network topologies

## VII. Conclusion

In this paper, we propose a novel AISC deployment model tailored for the dynamic UMEC network. Based on this model, we introduce the DDAQ-HGNN algorithm, which can learn structure-aware features from heterogeneous graph representation with multiple node and edge types and utilizes attention mechanisms to achieve intelligent AISC deployment in the dynamic UMEC network. Extensive experimental results demonstrate the effectiveness of DDAQ-HGNN in collaboratively optimizing AISC completion time, network load balancing, AISC completion rate and energy consumption. Compared with the baseline algorithms, DDAQ-HGNN exhibits superior adaptability and generalization capability, consistently achieving optimal performance under different network topologies and different UMEC network topology change probabilities.

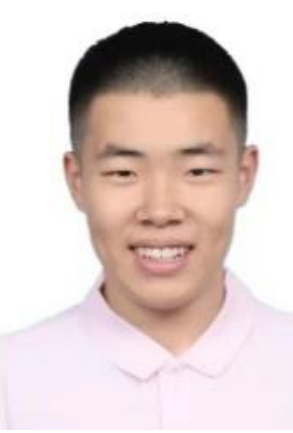

**Hanzhi Chang** (Student Member, IEEE) received his B.S. degree from the Department of Cyber Science and Engineering, University of International Relations, Beijing, China, in 2023. He is currently pursuing for his M.S. degree in the Department of Cyber Science and Engineering, University of International Relations, Beijing, China. His research interests include network function virtualization, network resource orchestration and management, and reinforcement learning algorithms.

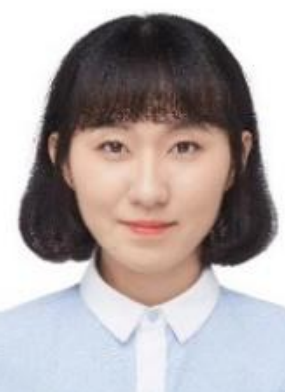

**Jing Bai** received the PhD degree in cyberspace security from Beijing Jiaotong University in 2023. She is currently a Lecturer at School of Cyber Science and Engineering, University of International Relations. Her interests include software trustworthiness analysis and cloud security.

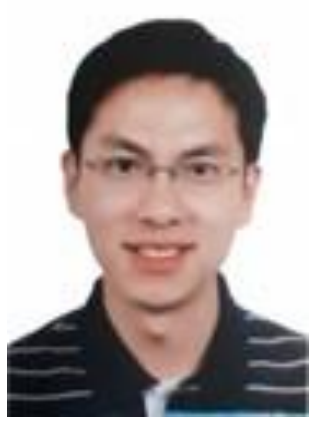

**Xin Tang** received the PhD degree in computer science from Beijing University of Posts and Telecommunications, Beijing, China, in 2015. He worked as a post-doctoral fellow in Department of Electronic Engineering at Tsinghua University, Beijing, China, from 2015 to 2017. He is currently an associate professor with the School of Cyber Science and Engineering, University of International Relations, Beijing, China. His current research interests are focused on secure cloud computing and reversible data hiding.

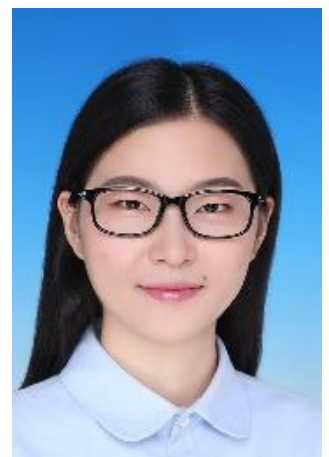

**Xiaomei Liu** received the Master's degree in computer technology from University of Chinese Academy of Sciences in 2018. She is currently an Engineer at School of Cyber Science and Engineering, University of International Relations. Her interests include cloud-edge computing and network security.